# Equation of state for tungsten predicted by ensemble theory

Yue-Yue Tian[1, 2], Bo-Yuan Ning[1, 2, 3], X.-D. Xiang[3], Hui-Fen Zhang[1, 2, 4*], Xi-Jing Ning[1, 2*]

[1]Institute of Modern Physics, Fudan University, Shanghai 200433, People's Republic of China

[2]Applied Ion Beam Physics Laboratory, Fudan University, Shanghai 200433, People's Republic of China

[3]Department of Materials Science and Engineering, Southern University of Science and Technology, Shenzhen 518055, People's Republic of China

[4]Hongzhiwei Technology (Shanghai) Co., Ltd., Shanghai 201206, People's Republic of China

**Abstract** Equation of state (EOS) for *bcc* tungsten at 300 K (or 3000 K) up to 1000 GPa (or 300 GPa) was predicted for the first time by solving the partition function via a direct integral approach (DIA) with *ab initio* calculations of the atoms' interactions. Compared with available experiments under static compressions up to 150 GPa (or 35 GPa) for room temperature (or 1673 K), all the calculated results are within the experimental uncertainty achieved very recently. Furthermore, the same procedure was performed to investigate the shock wave experiments on the EOS up to 400 GPa and 10000 K, and the calculated average pressure deviates the experimental measurements by only 2.0%. These facts suggest that the other calculated results of DIA for the EOS are reliable, and DIA as a universal method without any artificial parameters could be widely applied to predict EOS of various materials under various conditions.

## 1. Introduction

Tungsten (W), crystallizing in a body centered cubic (*bcc*) structure under ambient condition, is one of the most popular refractory materials for its ultrahigh melting point (3683 K). *Ab initio* total energy and temperature dependent phonon model calculations showed that *bcc* structural W will not change until 1.2 TPa [1] and the phase transition pressure point rises with increasing of temperature [2]. On account of the ultrahigh thermal and structural stability, metal W is regarded as an ideal pressure calibrant in static compression experiments, in which precision of pressure standards is crucially important to the high pressures research. So comprehensive and accurate command of equation of state (EOS), i.e. dependence of the volume ($V$) of W on pressure ($P$) and (or) temperature ($T$), is quite essential.

Up to date, accurate experimental room temperature measurements could be available up to megabar [3], beyond which the measured pressure may be unauthentic. On the other hand, performance of experiments is usually getting harder and harder with elevation of temperature that *P-V-T* measurements are always conducted at room temperature and up to relatively high pressure or at high temperature and lower pressure. For element W, the highest pressure conducted under room temperature is up to 150 GPa [4], while high temperature and simultaneous high pressure experiments were just reported by Dubrovinsky *et al.* up to 46.6 GPa, 1250 K [5], Litasov *et al.* up to 33.5 GPa, 1673 K [6] and Qi *et al.* up to 10.5 GPa and 1073 K [7]. These data are far insufficient for real applications.

In order to achieve simultaneous high temperatures and high pressures *P-V-T* knowledge, various empirical formulations [8-12] have been developed. With the empirical equation, experimental data are pre-requisitely fitted so that any *P-V-T* values within the experimental conditions and beyond could be obtained by making corresponding interpolations and extrapolations, respectively. Whereas, empirical EOS were always built on a very limited physics basis that the extrapolated results are always disparate using different empirical equations [10]. More notably, although with one empirical equation, extrapolated results would be diverse under different experiments covering different ranges of compressions [13]. So searching for a universal EOS is the absolutely required thing all the time.

Statistical physics brings a promising prospect for obtaining EOS of condensed matters without any empirical parameters as long as the partition function (PF) or free energy (FE) can be calculated precisely. Unfortunately, solving the PF involving a 3$N$-fold configurational integral far exceeds the capability of current computer technology even with the empirical potential describing the interactions of atoms, not to say *ab initio* computations, which severely hinders the practical applications of ensemble theory since it was built in the beginning of the twentieth century.

Phonon model based on quasi-harmonic approximation (QHA) supplies a possible route to PF and has been used to calculate EOS for many crystals [14-17]. Whereas, as is known that QHA is just available to small vibration situations, phonon model always works well at lower temperatures, but not at high ones. In fact, an elaborate work by Gong *et al.* [18] confirmed that the accuracy of QHA is not only dependent on temperature but also on the atomic volume of specific crystal that QHA manifests well just for the systems of atomic volume smaller than 22 Å$^3$ and the accuracy will get lower and lower gradually with increasing of the atomic volume.

For the above predicament, following the formalism of Helmholtz FE from QHA, $F_{\text{QHA}}(V, T)$ = $E_0(V) + F_{\text{ion}}(V, T)$, where $E_0$ is the total energy at 0 K with ions frozen at equilibrium positions and $F_{\text{ion}}(V, T) = \sum_{i=1}^{3N}(\frac{1}{2}\hbar\omega_i + k_B T ln(1 - e^{-\frac{\hbar\omega_i}{k_B T}}))$ is the vibrational contribution of ions to FE with $\omega_i$ the $i$th phonon mode of the lattice, scholars steer clear of the high-dimension configurational integral in PF but pay attention to make modifications for $F_{\text{ion}}(V, T)$ to improve the limitation of QHA. Xiang *et al.* [19] added a term $F_{\text{an}}$ separately to $F_{\text{QHA}}(V, T)$ as the reflection of an-harmonic vibration of ions, under which the calculated Hugoniot EOS of W up to ~ 250 GPa agree with the shock wave experiments [20, 21]. Whereas, the derivation of $F_{\text{an}}$ is on the assumption that internal energy of solids is independent of $T$, which is apparently untenable at finite temperatures. Another fashionable way to process $F_{\text{ion}}(V, T)$ in the late 1990s and early twenty-first century is the particle in a cell (PIC) model, in which interatomic correlations and diffusions are ignored, i.e. the vibrations of atoms are independent with each other, and the translational degrees of freedom of atoms are completely separable [22-24], so that the vibrational contribution of ions to the PF takes the form of $Z_{\text{ion}} = (\frac{2\pi m k_B T}{h^2})^{3N/2}[\int e^{-(U(r)-U_0)/k_B T} d\mathbf{r}]^N$, where $U_0$ and $U(\mathbf{r})$ is the potential energy of a system with all the $N$ atoms fixed on their ideal lattice sites and one wanderer atom displaced by the radius vector **r** from its equilibrium position, respectively, and the integral is over the Wigner-Seitz cell centered on the wanderer atom. PIC is essentially an an-harmonic Einstein model and simplifies the 3$N$-fold integral in PF into a triple one, but which is still a nearly impossible thing with accurate *ab initio* calculations. To solve this problem, $U(\mathbf{r})$ was usually constructed by empirical potentials, such as the tight binding total energy method by Wasserman *et al.* [22], and pair-wise potential by Xiang *et al.* [25, 26]. Nevertheless, as is well known that empirical potentials could not depict the real interactions between atoms of materials, PIC is still not a reliable method. It is worth mentioning that Wang substituted $U(\mathbf{r})$ by a classical mean-field potential [27-32] (MFP) dependent just on $\mathbf{r}$ leading to a one-dimensional integration of PF. Although MFP was calculated from first principles, it is too simple that cannot suit to various elements and solids. Afterwards, Wang promoted another formalism of potential [33], in which there exists adjustable parameter. Obviously, MFP is also not a universal method to calculate the thermodynamic properties of condensed matters.

Different from all the previous methodologies, a direct integral approach (DIA) directly to the PF without any empirical or artificial parameters was put forward very recently by Ning *et al.* [34], who reinterpreted the high-dimension configurational integral of PF as an effective volume, leading to ultrahigh computation efficiency even with *ab initio* calculating the total potential of materials. The mathematical derivation for DIA and its validation has been rigorously demonstrated in ref. [34] and would be left out here. DIA works universally, operates very simply with ultrahigh precision and has been successfully applied to reproduce the EOS of solid copper [34], argon [35], 2-D materials [36] and phase transition of crystal vanadium [37], zirconium [38] and aluminum [39]. Especially, a recent calculation for EOS of iridium based on DIA [40] agrees very well with experiments and is much better than that from MFP and QHA.

In the present work, DIA was used to calculate PF of crystal W and the further derived EOS were directly compared with previous experiments. Available room temperature (and high temperatures (< 1673 K)) isothermal static compression experimental EOS up to 150 GPa (and 35 GPa) and Shock wave Hugoniot EOS up to 5400 K and 260 GPa were totally reproduced. Specific comparisons show that QHA always underestimate the pressure of W and it manifests worse than DIA even under room temperature.

## 2. DIA calculation method

For a crystal containing $N$ atoms confined within volume $V$ at temperature $T$, the atoms are regarded as $N$ point particles of atomic mass $m$ with cartesian coordinate $\boldsymbol{q}^N = \{\boldsymbol{q}_1, \boldsymbol{q}_2, \cdots \boldsymbol{q}_N\}$, and the total potential energy, $U(\boldsymbol{q}^N)$, as the function of $\boldsymbol{q}^N$ can be computed by quantum mechanics, i.e., for a given set of $\boldsymbol{q}^N$, the total potential energy $U(\boldsymbol{q}^N)$ concerned with the motions of electrons in the field of the nucleus fixed at the lattice sites is calculated by quantum mechanics. With knowledge of $U(\boldsymbol{q}^N)$, the PF of the system reads

$$Z = \frac{1}{N!}\left(\frac{2\pi m}{\beta h^2}\right)^{\frac{3}{2}N} \int d\boldsymbol{q}^N e^{-\beta U(\boldsymbol{q}^N)} = \frac{1}{N!}\left(\frac{2\pi m}{\beta h^2}\right)^{\frac{3}{2}N} Q, \qquad (1)$$

where $h$ is the Planck constant and $\beta = 1/k_B T$ with $k_B$ is the Boltzmann constant. If the configurational integral $Q = \int d\boldsymbol{q}^N e^{-\beta U(\boldsymbol{q}^N)}$ is solved, the pressure $P$, Helmholtz FE and the internal energy $E$ can be calculated via

$$P = \frac{1}{\beta}\frac{\partial \ln Z}{\partial V} \qquad (2)$$

$$F = -\frac{1}{\beta}\ln Z \qquad (3)$$

$$E = -\frac{\partial \ln Z}{\partial \beta} \qquad (4)$$

According to DIA [34], for a single component crystal with $N$ atoms placed in their lattice sites $\boldsymbol{Q}^N$ and with the total potential energy $U_0(\boldsymbol{Q}^N)$, we firstly introduce a transformation,

$$\boldsymbol{q}'^N = \boldsymbol{q}^N - \boldsymbol{Q}^N, \; U'(\boldsymbol{q}'^N) = U(\boldsymbol{q}'^N) - U_0(\boldsymbol{Q}^N), \qquad (5)$$

where $\boldsymbol{q}'^N$ and $U'(\boldsymbol{q}'^N)$ represent the displacements of atoms away from their lattice positions and the corresponding potential energy differences with respect to the $U_0(\boldsymbol{Q}^N)$, and the configurational integral $Q$ can then be expressed in a way of one-fold integral as

$$Q = e^{\beta U_0}\left[\int e^{-\beta U'\left(q'_{i_{x,y,x}}\right)}dq'_{i_{x,y,x}}\right]^{3N} \qquad (6)$$

Where $q'_{i_{x,y,x}}$ denotes the distance of the $i$th atom moving along the $x$ (or $y$, $z$) direction relative to its lattice site while the other two degrees of freedom and all the other atoms are kept fixed.

As shown in Fig. 1(a), a $3 \times 3 \times 3$ (54 atoms) supercell of *bcc* W with lattice constant $a_0 = 3.172$ Å was used in the present work. According to Eq. (6), an arbitrary atom (in the red dashed circle) was selected and moved away gradually along the <111> direction by a step of 0.03√3 Å until the total potential energy increases by about 12 eV, which is enough for the contribution to the integration in Eq. (6) up to 10000 K. At each step of the moved atom, the total potential energy of the system is calculated by the density functional theory (DFT) and spline interpolation algorithm [41] is used to smooth $U'(x')$ (Fig. 1(b)). Changing the volume of W by enlarging or shrinking the lattice constant, then with the same process of moving one atom step by step and

$U'(x')$ curve under a series of volume can be obtained, which are shown in Fig. 1(b). Finally, EOS of crystal W could be obtained on the basis of formula (1), (2) and (6).

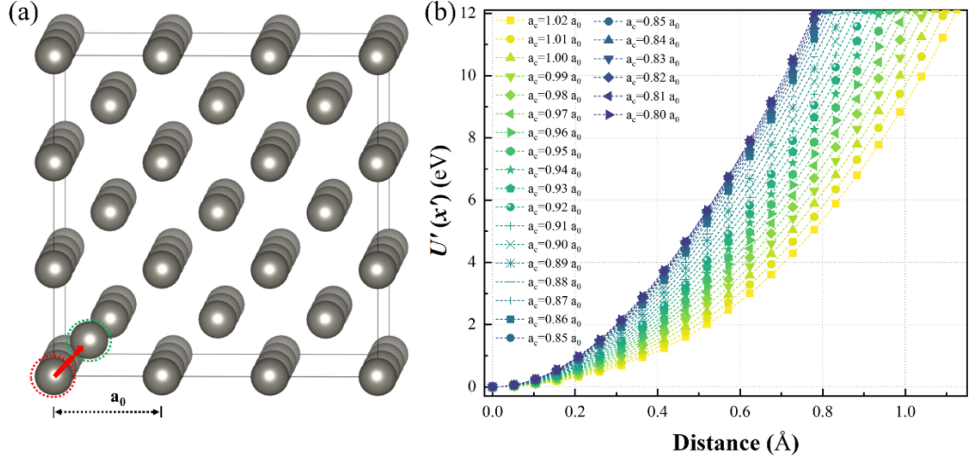

Fig.1 (a) A $3 \times 3 \times 3$ supercell of *bcc* W with lattice constant $a_0 = 3.172$ Å, in which an arbitrary atom in the red dashed circle is moved along the arrow direction step by step and arrives at its final position represented by the atom in the green circle; (b) the calculated $U'(x')$ under different supercell volume with lattice constant $a_c$ changed from 0.80 $a_0$ to 1.02 $a_0$.

DFT calculations in the present work were implemented in Vienna Ab initio Simulation Package [42, 43], in which Projector Augmented Wave pseudopotential [44, 45] and Perdew Burke Ernzerhf of General Gradient Approximation [46] were selected to describe the electron-ion interactions and electron-electron change-correlation functional with 6 valence electrons ($6s^2 5d^4$) considered. The plane-wave cut-off energy was 400 eV and a Γ-centered $5 \times 5 \times 5$ uniform k-mesh grid is set to sample the Brillouin zone by the Monkhorst-Pack scheme. Electron self-consistent tolerance for total energy is $2 \times 10^{-6}$ eV.

3. Results and discussions

3.1 Room temperature isothermal EOS

Based on Eq. (1), (2) and (6), isothermal *P-V* curve under 298 K is shown as the red dashed line in Fig. 2(a). To evaluate the validation of DIA, direct comparisons for the calculated pressure ($P_{DIA}$) with three latest experimental measurements ($P_{Exp}$) under same value of $V/V_0$ are conducted and exhibited as the red squares in Fig. 2(b). For the experiment by Qi *et al*. [7] containing seven experimental *P-V* points, deviations of $P_{DIA}$ distribute evenly from $P_{Exp}$ and all the | $P_{Exp} - P_{DIA}$ | values are within the uncertainty of the experimental measurement (0.2 GPa) except for one,

0.25 GPa, which should result from the offset of spline algorithm used in the present work to smooth the *P-V* data. The deviations of $P_{DIA}$ from experimental pressure of Litasov *et al.* [6] distribute also evenly in the whole pressure range (0 ~ 31 GPa) and the largest value of $P_{Exp} - P_{DIA}$ is 1.15 GPa. Averagely, the difference between calculated pressure from DIA and the experiment of ref. [6] is 0.44 GPa, which is slightly larger than the uncertainty, 0.3 GPa, of the used MgO pressure scale in ref. [6]. In the static compression experiment of Dewaele *et al.* [4], ruby was used as the pressure scale with accuracy smaller than 2% up to 55 GPa, within which the average value of $|P_{Exp} - P_{DIA}|$ is 0.6 GPa. Beyond 55 GPa, the calculated $P_{DIA}$ gradually deviates the experimentally measured pressure with increase of compression and the result of $P_{Exp} - P_{DIA}$ reaches -10.6 GPa at the maximum experimental pressure 144 GPa. Several previous works indicated that the ruby pressure scale could underestimate pressure by ~ 10 GPa at 150 GPa [47-49], which expectedly indicated that calculations of DIA are indeed reliable.

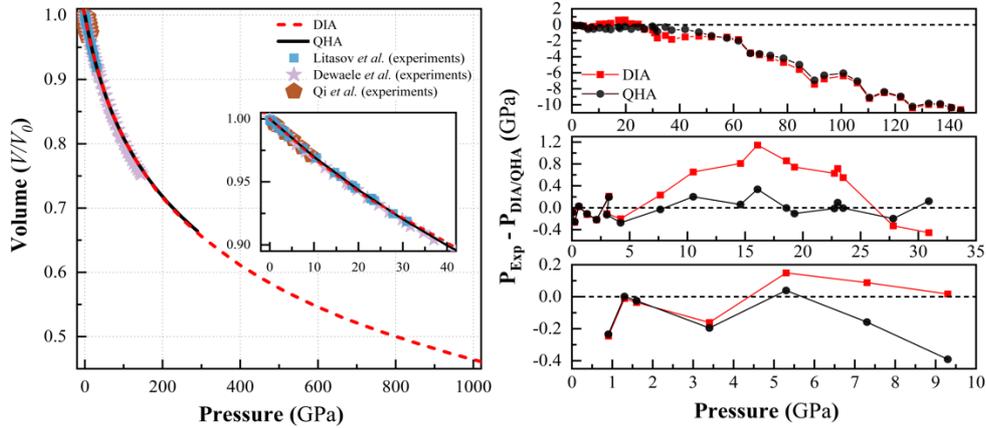

Fig.2 Dependence of relative volume of W $V/V_0$ on the pressure under room temperature. $V_0$ is the volume at one atmospheric pressure. Red dashed and black solid line are the calculated results from DIA and quasi-harmonic approximation (QHA) by Xiang *et al.* [19], respectively. Blue squares, purple stars and brown pentagons represent experimental measurements from Litasov *et al.* [6], Dewaele *et al.* [4] and Qi *et al.* [7]; (b) Difference of calculated pressure from DIA ($P_{DIA}$) (red squares) or QHA ($P_{QHA}$) by Xiang *et al.* [19] (black circle) with the measured one ($P_{Exp}$) from experiment of Dewaele *et al.* [4] (upper one), Litasov *et al.* [6] (middle one), and Qi *et al.* [7] (bottom one) under the same value of $V/V_0$.

Room temperature isothermal EOS of W has been calculated by Xiang *et al.* [19] under quasi-harmonic approximation (QHA), which is also presented in Fig. 2(a) as the black solid line. It seems that QHA works well in this case, but it is actually worse than DIA. To demonstrate this, direct comparisons of the calculated pressures from QHA, $P_{QHA}$, with the experiments of ref. [7], [6]and [4] were also conducted and the results are displayed as the black circle dots in Fig. 2(b). For experiment of ref. [6], calculated pressure from QHA is indeed closer to the experiment than that of DIA shown as the middle graph in Fig. 2(b). But for the latest experiment [7], all the $P_{Exp}$-$P_{QHA}$ values are negative except for one, which is only slightly larger than the corresponding experimental measured result by 0.001 GPa. What's more, calculated pressures from QHA are smaller than the experimentally measured results in the whole pressure range (0 ~ 150 GPa) of experiment ref. [4]. From the above discussions, QHA may generally underestimate pressures for the calculations of W.

## 3.2 High temperature isothermal EOS

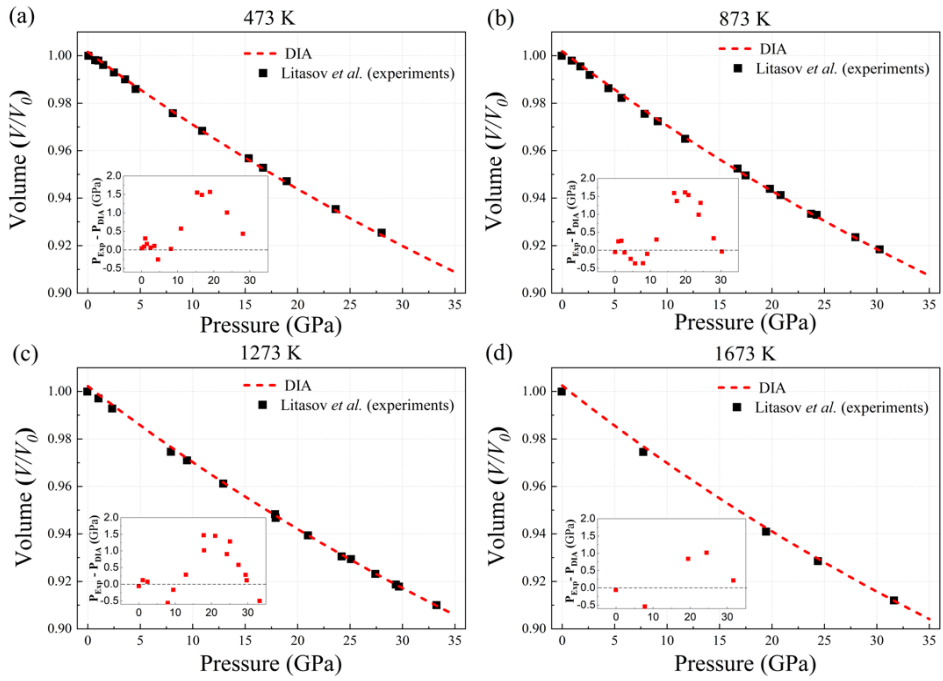

Fig.3 Isothermal dependence of relative volume $V/V_0$ on the pressure under 473 K (a), 873 K (b), 1273 K (c) and 1673 K (d) with $V_0$ the one atmospheric volume at the respective temperature. Red dashed lines are the calculated results from DIA and the black squares represent measured values of experiment ref. [6]. Illustrations are the deviation of the calculated pressure by DIA from the experimental one under same $V/V_0$.

With a simple operation in DIA as the second section expressed, isothermal EOS of W under any high temperature could be easily obtained just via substituting the temperature $T$ in Eqs. (1-3) and (6) by specific value. To make a direct check for the reliability of DIA under high temperatures, experiment by Litasov *et al.* [6] was taken because their maximum experimental temperature (1673 K) is the highest one among the available static compression experiments. Isothermal $P$-$V$ curves for W under 473, 873, 1273 and 1673 K by DIA is shown as red dashed line in Fig. 3(a), (b), (c) and (d), respectively and the black scattered squares are the experimental measurements in ref. [6]. From the illustrations of Fig. 3, we can see that deviations of the calculated pressure, $P_{DIA}$, from the experimental $P_{Exp}$ under the same $V/V_0$ distributed evenly within the whole experimental conditions in all the four cases, and the average value of ∣$P_{Exp}$ - $P_{DIA}$∣, 0.55, 0.64, 0.59, 0.54 GPa, respectively, are quite comparable with that of room temperature (0.44 GPa), indicating the accuracy of DIA would not degrade with the increment of temperature.

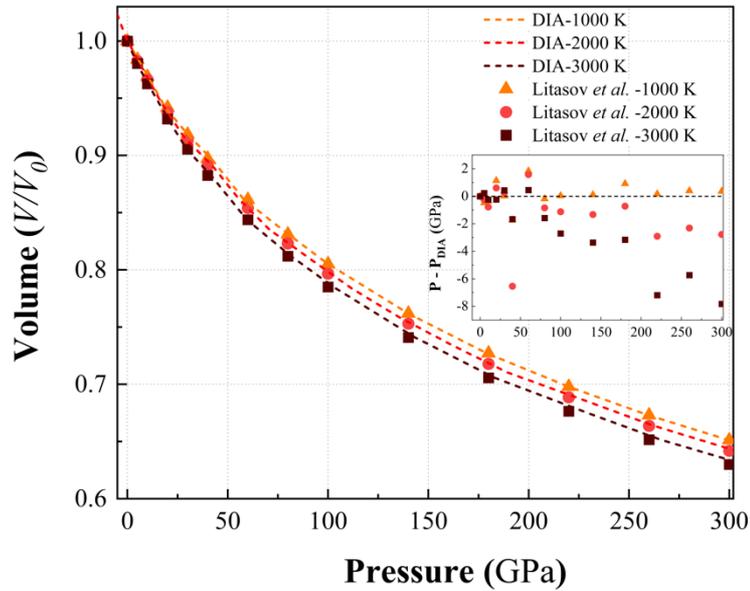

Fig. 4 Isothermal dependence of relative volume $V/V_0$ on the pressure under 1000 K, 2000 K and 3000 K from DIA (dashed lines) and KE EOS (scatter points) with $V_0$ the one atmospheric volume at the respective temperature.

In work of ref. [6], empirical Kunc-Einstein (KE) equation was used to make high temperature and high pressure EOS predictions. After the parameters in KE equation were obtained, isothermal EOS of W from ambient condition up to 3000 K and 300 GPa were presented. To have a representative comparison, results from KE equation under 1000, 2000 and 3000 K together with that calculated from DIA were presented in Fig. 4, and the pressure difference between them under the same $V/V_0$ is shown in the illustration. Under 1000 K, calculated pressure from DIA ($P_{DIA}$) evenly deviated the pressure from KE equation (P) and the average value of $|P - P_{DIA}|$ is 0.55 GPa. While in the situation of 2000 K, it is clear that above 80 GPa, departure of $P_{DIA}$ from P is more and more with the increase of compression that the average value of $|P-P_{DIA}|$ is 1.57 GPa, which is about three times larger than that within the experimental conditions (< 1673 K) of ref. [6]. What's more, the difference between $P_{DIA}$ and P under 3000 K is much larger and the average value of $|P-P_{DIA}|$ increases to 2.49 GPa. As mentioned in the first section of the present work, results from empirical EOS are strongly dependent on the accuracy of the based experiments and the pressure measurements beyond ~ 100 GPa are always unauthentic. On the other hand, from the physical background of DIA [34], its accuracy would get higher with gradually increased density of solids. So the above deviation of $P_{DIA}$ from P above 80 GPa under 2000 and 3000 K should account for the low accuracy of the used experimental data. This could also be explained by the following section that Hugoniot EOS of W from DIA agrees very well the shock wave experiments until 5400 K and 260 GPa.

### 3.3 Hugoniot EOS

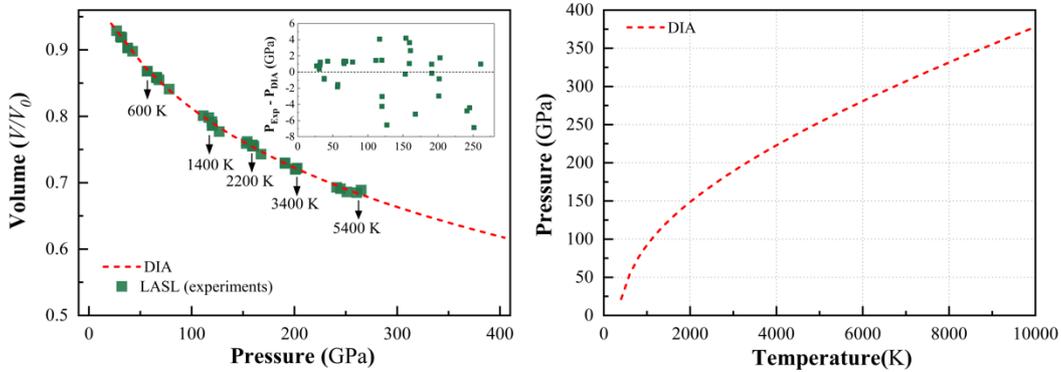

Fig. 5 Hugoniot EOS of W calculated from DIA (red dashed line) and shock wave experiment (green scattered points) [20]. The temperature marker in the left graph is the calculated result from DIA. The right *P-T* curve corresponds to the left Hugnoniot *P-V* line from DIA.

Unlike the pressure determining in static compression experiments that are basically dependent on the accuracy of the used secondary pressure scale, pressures in shock wave experiments could be directly measured leading to smaller system errors. In addition, experimental pressure and temperature could reach a higher scope in shock wave experiments, which supplies valuable data to further examine the accuracy of DIA.

To compare with shock wave experiments, Hugoniot pressure $P_H$ versus volume $V$ of W should be calculated first. According to the Rankine-Hugoniot equation, $P_H (V_0 - V) = 2(E_H - E_0)$, where $E_0$ and $V_0$ are respectively the internal energy and volume at ambient condition, under temperature $T$, $P_H$ and $E_H$ are calculated based on Eq. (2) and (4) from DIA with variable $V$ until the Hugoniot equation is satisfied. Changing $T$, another group of $P_H$-$V$-$T$ could be obtained with similar procedure. With this process Hugoniot EOS of W is presented in Fig. 5. To compare our results with shock wave experiment of ref. [20], difference of pressure calculated by DIA and experimental measurement is computed and presented in the illustration of Fig.5 (a), which clearly shows that deviations of $P_{DIA}$ distribute evenly in the whole pressure range (0 ~ 260 GPa), and the average value of ∣$P_{exp}$-$P_{DIA}$∣ is only 2.2 GPa, indicating the precision of DIA is high enough up to 5400 K and 260 GPa. Calculated Hugoniot pressure in the present work is just up to 400 GPa and temperature up to 10000 K, beyond which phase transition of *bcc* W was indicated in previous shock wave experiment [21].

4. Conclusion

The rigorous ensemble theory is applied, for the first time, to calculate the isothermal EOS of *bcc* W at 300 K (or 3000 K) up to 1000 GPa (or 300 GPa) and Hugoniot EOS up to 400 GPa and 10000 K by DIA solving the PF without any empirical or artificial parameters. Available static compression experimental EOS under room temperature (and high temperatures) up to 150 GPa (and 35 GPa) and shock wave Hugoniot EOS up to 5400 K and 260 GPa were completely reproduced, indicating other predicted EOS results for *bcc* W are predicted to be reliable. DIA is universal to various materials with ultrahigh efficiency and could be used to predict thermodynamic properties in various conditions. It is convincible that DIA would find its essential applications in the thermodynamic field in the future.

**Acknowledgement**

Part of the computational tasks were conducted in HPC platform supported by The Major Science and Technology Infrastructure Project of Material Genome Big-science Facilities Platform supported by Municipal Development and Reform Commission of Shenzhen.